# Self-trapping and splitting of bright vector solitons under inhomogeneous defocusing nonlinearities

Yaroslav V. Kartashov,[1] Victor A. Vysloukh,[1] Lluis Torner,[1] and Boris A. Malomed[2,1]

[1]*ICFO-Institut de Ciencies Fotoniques, and Universitat Politecnica de Catalunya, Mediterranean Technology Park, 08860 Castelldefels (Barcelona), Spain*
[2]*Department of Physical Electronics, School of Electrical Engineering, Faculty of Engineering, Tel Aviv University, Tel Aviv 69978, Israel*

We show that bimodal systems with a spatially nonuniform defocusing cubic nonlinearity, whose strength grows toward the periphery, can support stable two-component solitons. For a sufficiently strong XPM interaction, vector solitons with overlapping components become unstable, while stable families of solitons with spatially separated components emerge. Stable complexes with separated components may be composed not only of fundamental solitons, but of multipoles too.

A commonly adopted principle, which has been thoroughly tested in nonlinear optics, states that a focusing nonlinearity is necessary to support bright solitons in uniform media [1]. The situation changes in the presence of linear lattice potentials, which allow formation of gap solitons even in defocusing media [2]. Recent advances in the fabrication of artificial optical media allow creation of various nonlinearity landscapes, which may strongly affect the soliton dynamics, as the respective pseudo-potential depends on the light intensity distribution [3]. Still, in contrast to their linear counterparts, nonlinear lattices formed by a periodic modulation of defocusing nonlinearity do not support bright solitons.

However, the situation may be drastically different if the strength of the defocusing nonlinearity grows at $r \to \infty$, where $r$ is the transverse coordinate. It has been demonstrated recently that such nonlinearity patterns can support stable bright solitons [4,5]. Such a counterintuitive finding is based on the fact that the corresponding equation for the field amplitude is non-linearizable at $r \to \infty$, a property that drastically affects decaying tails of solitons, invalidating the commonly known proof of the nonexistence of bright solitons under defocusing nonlinearities.

In this Letter we show that the variety of bright solitons supported by the rising defocusing nonlinearity may be vastly expanded in two-component systems. Such systems support stable vector solitons with fundamental and multipole components. We show that the increase of the XPM (cross-phase-modulation) strength between the components causes a *splitting transition* from solitons with spatially overlapping components to ones with separated components. This transition was analyzed previously only in free space [6] or in linear potentials [7,8], but it was never demonstrated in inhomogeneous nonlinear landscapes. Furthermore, such a splitting transition is obtained here for the first time for multipole states. Our results also apply to matter-wave solitons in binary Bose-Einstein condensates (BECs).

The spatial evolution of amplitudes $q_{1,2}$ of vector optical beams, or the temporal evolution of wave functions of the binary BEC, in media with the spatially inhomogeneous cubic nonlinearity obeys the system of coupled nonlinear Schrödinger equations [9-11]:

$$i\frac{\partial q_{1,2}}{\partial \xi} = -\frac{1}{2}\frac{\partial^2 q_{1,2}}{\partial \eta^2} + \sigma(\eta)q_{1,2}(|q_{1,2}|^2 + C|q_{2,1}|^2), \qquad (1)$$

Here $\xi$ is the propagation distance (or time in BEC), $\eta$ is the transverse coordinate, $C$ is the XPM strength that depends on polarization state, wavelength, and mutual coherence of beams [1], and function $\sigma(\eta) > 1$ describes spatial profile of the nonlinearity. In BEC the interspecies repulsion strength can be controlled via Feshbach resonance in nonuniform external fields [3,12]. In optical media the transverse modulation of the nonlinearity can be created in a variety of settings [3]. Thus, doping of photovoltaic materials, such as $LiNbO_3$, with Cu or Fe can be used to enhance the local defocusing nonlinearity [13-15]. The nonlinear refractive index change in $LiNbO_3$ may reach the level of $\sim 10^{-3}$, and it varies by several orders of magnitude depending on the local concentration of dopants that can be made spatially nonuniform using indiffusion of the dopant layer with a varying width (that, together with the indiffusion time determines the dopant concentration) into the surface of $LiNbO_3$ crystal [15]. In addition, in externally biased photorefractive media, a nonuniform background illumination with intensity $I_{bg}(\eta)$ leads to a nonlinear contribution to the refractive index, $\sim E_0[1 - I/I_{bg}(\eta)]$, for $I \ll I_{bg}$ ($I$ is the intensity of the probe beam, and $E_0$ is the biasing field), which induces the nonlinearity modulation. Here we consider the modulation profile with $\sigma(\eta) = \exp(\alpha \eta^2)$, and fix $\alpha = 1$ by rescaling.

Soliton solutions $q_{1,2}(\eta, \xi) = w_{1,2}(\eta)\exp(ib_{1,2}\xi)$ of Eq. (1) are characterized by the propagation constants $b_{1,2}$ of the two components, the total and partial energy flows $U = U_1 + U_2 = \int_{-\infty}^{\infty}(|q_1|^2 + |q_2|^2)d\eta$, and by the energy sharing ratio, $S_{1,2} \equiv U_{1,2}/U$. Examples of vector solitons whose components fully overlap, being centered at $\eta = 0$, are displayed in Figs. 1(a),(b). As it follows from the stationary version of Eq. (1) taken at the inflection points, where $\partial^2 q_{1,2}/\partial \eta^2 = 0$, the bright solitons may exist only for $b_1, b_2 < 0$. Such solitons exist too with $b_1 \neq b_2$ (hence the two components may carry different energy flows) in a finite interval, $b_2^{low} \leq b_2 \leq b_2^{upp}$, for fixed $b_1$ and $C$. The existence domain for the solitons with overlapping components shrinks at $C = 1$ and expands with increase of $C$, see Fig. 2(c).

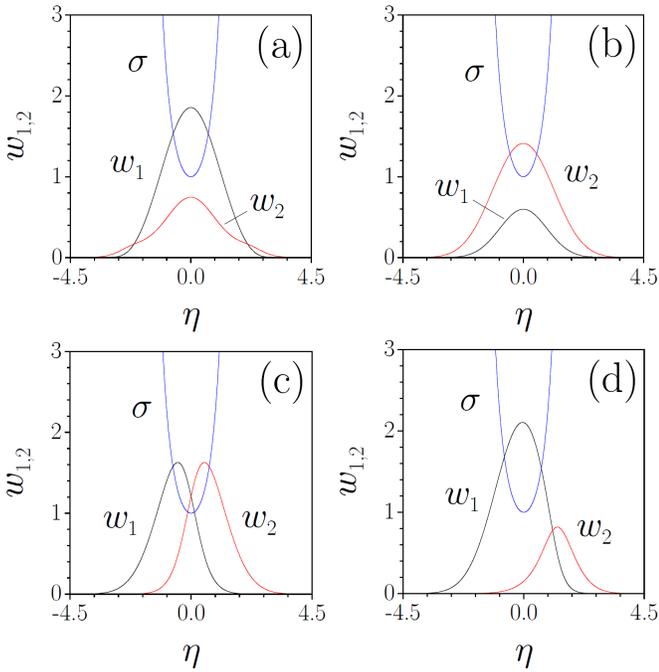

Figure 1. Fundamental vector solitons with overlapping components, for $b_2 = -8$ (a), $b_2 = -3.1$ (b), and with separated ones, for, $b_2 = -5$ (c), $b_2 = -7.2$ (d). In all the cases, $b_1 = -5$ and $C = 2$. The soliton in (a) is unstable, while the ones in (b)-(d) are stable.

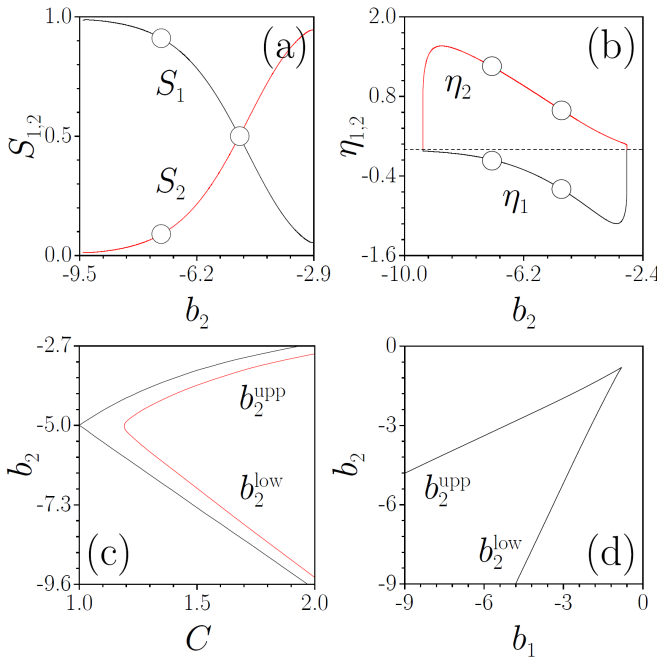

Figure 2. The energy-sharing ratio (a), and coordinates of centers of the two components (b), versus $b_2$ at $b_1 = -5$ and $C = 2$. Circles correspond to the solitons in Figs. 1(c) and 1(d). (c) The existence domains in the plane of $(C, b_2)$ for vector solitons with overlapping and separated components (black and red curves, respectively) at $b_1 = -5$. (d) The existence domains of vector solitons with separated components in the $(b_1, b_2)$ plane at $C = 2$.

The solitons with overlapping components, that are stable at $C \simeq 1$, become unstable at larger values of $C$ (the stability was analyzed by finding the eigenvalue spectrum from linearized equations for small perturbations). Simultaneously, a new type of stationary vector solitons emerges, with spatially separated components [see Figs. 1(c) and 1(d)]. These solitons exist precisely in the parameter region where the ones with overlapping components are unstable, and their existence domain is embedded into the broader domain of existence for the solitons with overlapping components [Fig. 2(c)]. At $b_1 = b_2$, both components of the soliton with separated components carry equal energy flows [Fig. 1(b)], but the asymmetry between them becomes conspicuous at $b_1 \neq b_2$ [Fig. 1(c)], with the energy fraction carried by first component decreasing with the increase of $b_2$ [Fig. 2(a)].

Especially interesting is the dependence of the location of centers of two components on $b_2$. There exist values of $b_2$ at which the high-amplitude component is located almost at the center, while other low-amplitude component is shifted far to the periphery [Fig. 2(b)]. However, close to the existence borders, at $b_2 \to b_2^{\text{low}}, b_2^{\text{upp}}$, the weak component experiences considerable deformation and shifts to the center, so that the soliton resembles a scalar one centered at $\eta = 0$, with nearly all the energy concentrated in one component.

At fixed $b_1$ vector solitons with separated components are found only above a certain threshold value of the XPM coefficient, $C$. Thus, at $b_1 = -5$ the existence domain shrinks at $C_{\text{th}} \approx 1.19$ to point $b_2 = b_1$, at which the separation between the components vanishes [Fig. 2(c)]. Similarly, at fixed $C$ the existence domain in the plane of $(b_1, b_2)$ shrinks with the increase of $b_1$ [Fig. 2(d)]. Vector solitons with separated fundamental components are stable almost in their entire existence domain [for any separation between the components from Fig. 2(b)] except for very narrow regions near boundaries $b_2^{\text{low}}, b_2^{\text{upp}}$, not even visible on the scale of Fig. 2(c).

The transition from the solitons with overlapping components to the separated ones can be analyzed by means of the variational approximation. To this end, soliton solutions with mutually symmetric components are approximated by ansatz $q_{1,2} = A(1 \pm \kappa \eta - \eta^2/2) \exp(-\eta^2/2) \exp(ib\xi)$, where $2\kappa$ is the separation, and, up to order $\kappa^2$, the energy flow does not depend on $\kappa$. Substituting $q_{1,2}$ into the Hamiltonian of Eqs. (1) and expanding it up to $\kappa^2$, the transition from the overlapping components to separated ones occurs when the coefficient in front of $\kappa^2$ term vanishes. This gives $C_{\text{th}} = (b - 1/4)/(b + 3/4)$, where we used the prediction of variational approximation for the amplitude of soliton with overlapping components, $A^2 = -(b + 1/4)/(C + 1)$ [5]. For $b = -5$ analytics gives $C_{\text{th}} \approx 1.23$, while numerically found threshold is $C_{\text{th}} = 1.19$. Recall that the commonly known result, originating from the competition between SPM and XPM in the free space, is $C_{\text{th}} \equiv 1$ [6]. The deviation of $C_{\text{th}}$ from 1 in our case is due to inhomogeneous nonlinearity.

Bright vector solitons exist not only in nonlinearity landscapes $\sigma(\eta) = \exp(\alpha \eta^2)$, but also for much slower laws of nonlinearity growth, such as $\sigma(\eta) = 1 + |\eta|^\alpha$, with any $\alpha > 1$. Such landscapes support vector solitons whose tails decay as $\sim (1 + |\eta|^\alpha)^{-1/2}$ at $|\eta| \to \infty$, and those solitons also feature splitting transition. The splitting between components in algebraic solitons is of the same order as in the model with $\sigma(\eta) = \exp(\alpha \eta^2)$, but the width of algebraic solitons may be considerably larger. Even if nonlinearity $\sigma(\eta) = \exp(\alpha \eta^2)$ saturates (ceases to grow) at $|\eta| > \eta_0$ (which is expected in realistic settings), this results in appearance of oscillating tails with a small amplitude, $\sim \exp(-\alpha \eta_0^2/2)$, that is virtually unresolvable already for $\eta_0 \sim 5\alpha^{-1/2}$.

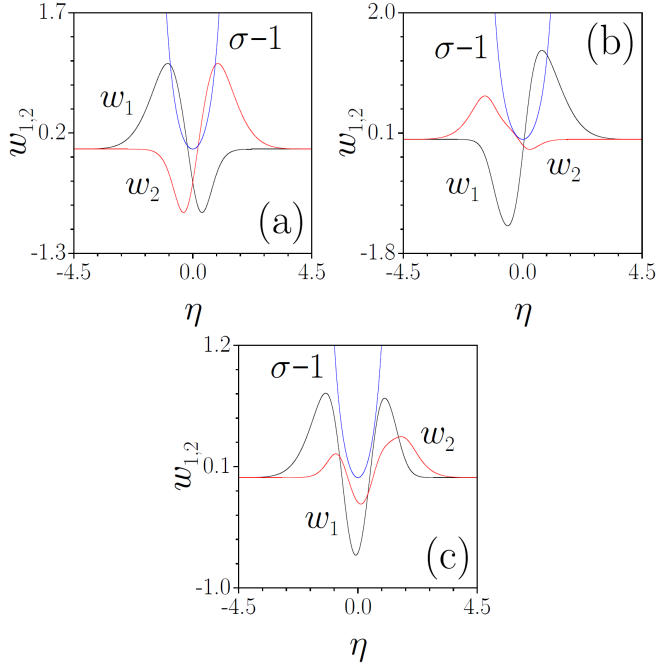

Figure 3. Dipole solitons with separated components at (a) $b_1=-5$, $b_2=-5$, (b) $b_1=-5$, $b_2=-7.4$ for $C=3$. (c) Tripole soliton with separated components at $b_1=-5$, $b_2=-6.5$, $C=3$. The solitons in (a) and (b) are stable, while the one in (c) is unstable.

Solitons with separated components can be built not only of fundamental modes, but also of multipoles. Examples of the solitons composed of dipole and tripole components are shown in Figs. 3(a)-3(c). Such composite states were not reported before, in either linear or nonlinear potentials. Multipole vector solitons feature symmetric separated components at $b_1=b_2$ [Fig. 3(a)], and strong asymmetry at $b_1\neq b_2$ [Fig. 3(b)]. The emergence of separated vector states is associated with the onset of the instability of overlapping dipole vector solitons. Multipole vector solitons feature qualitative properties similar to those of the fundamental solitons - for example, their first component vanishes with the increase of $b_2$ [Fig. 4(a)] (the total energy flow decreases with the decrease of $|b_2|$); they exist in a limited domain $b_2^{\text{low}} \leq b_2 \leq b_2^{\text{upp}}$ for fixed $b_1, C$; their existence domain shrinks with the decrease of $C$ [Fig. 4(b)] or increase of $b_1$ [Fig. 4(c)]. The existence domain for the multipole solitons is narrower than for their fundamental counterparts, and it shrinks with the increase of the number of poles in the soliton. For $b_1=-5$, vector solitons with separated components exist above the threshold XPM constant $C_{\text{th}} \approx 1.28$, which is higher than threshold $C_{\text{th}} \approx 1.19$ for existence of fundamental solitons.

Despite their complex structure, vector solitons with separated multipole components may be stable. The stability and instability domains for such solitons alternate with variation of $b_2$ (or, equivalently, of the separation between the components), as shown in Fig. 4(d). Unstable vector multipoles usually spontaneously transform into irregularly oscillating breathers. Stable perturbed two-component multipoles preserve their structures over indefinitely long distances.

In conclusion, we have found that the inhomogeneous defocusing nonlinearity, growing fast enough toward the periphery, supports stable bright vector solitons, composed of fundamental modes and multipoles. The system where the nonlinearity profile is different for two components (for example, when the nonlinearity is uniform in one component and modulated in the other) supports bound states of other types, such as dark-bright vector solitons.

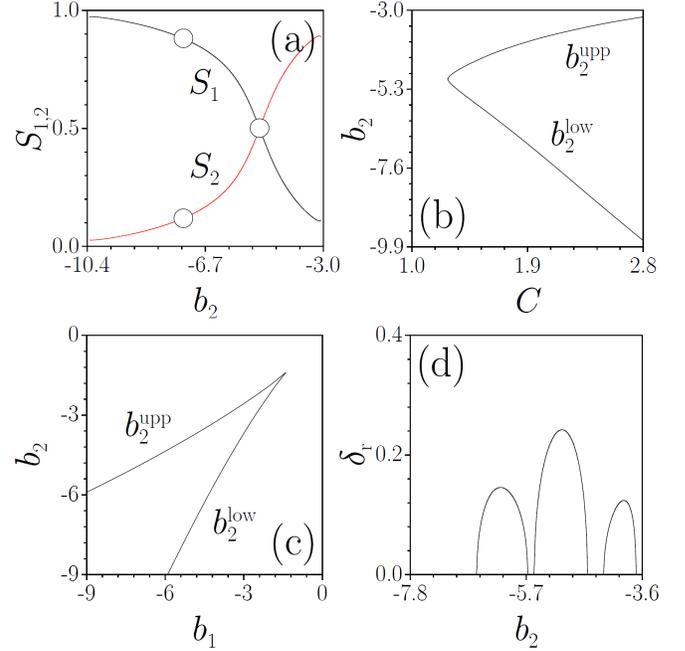

Figure 4. (a) The energy-sharing ratio versus $b_2$ at $b_1=-5$, $C=3$. Circles correspond to solitons in Figs. 3(a) and 3(b). The existence domains of dipoles with separated components in the plane of $(C, b_2)$ at $b_1=-5$ (b), and in the plane of $(b_1, b_2)$ at $C=2$ (c). (d) The real part of the perturbation growth rate versus $b_2$ at $b_1=-5$, $C=2.2$ for the dipole soliton with separated components.